\documentclass[12pt,onecolumn]{IEEEtran}

\usepackage{color}
\usepackage{amsmath}
\usepackage{amssymb}
\usepackage[dvips]{graphicx}
\usepackage{setspace}
\renewcommand{\baselinestretch}{1.8}

\newcommand{\rr}{{\bf r}}
\newcommand{\bb}{{\bf b}}
\newcommand{\nn}{{\bf n}}

\newcommand{\xx}{{\bf x}}
\newcommand{\yy}{{\bf y}}
\newcommand{\zz}{{\bf z}}
\newcommand{\ee}{{\bf e}}
\newcommand{\qq}{{\bf q}}
\newcommand{\pp}{{\bf p}}

\newcommand{\one}{{\bf 1}}
\newcommand{\zero}{{\bf 0}}

\newcommand{\ssT}{{\bf s}^T}
\newcommand{\ssV}{{\bf s}}

\newcommand{\xxx}{{\bf X}}
\newcommand{\rrr}{{\bf R}}
\newcommand{\sss}{{\bf S}}
\newcommand{\aaa}{{\bf A}}
\newcommand{\iii}{{\bf I}}

\newcommand{\ppp}{{\bf P}}

\newcommand{\mtL}{{\mathcal{L}}}

\newcommand{\MU}{\boldsymbol{\mu}}
\newcommand{\NU}{\boldsymbol{\nu}}

\begin{document}

\title{Optimal and Suboptimal Finger Selection Algorithms for MMSE Rake Receivers in Impulse Radio Ultra-Wideband Systems$^\textrm{\small{0}}$}
\author{Sinan Gezici$^{1}$, \textit{Member, IEEE}, Mung Chiang$^{2}$,
\textit{Member, IEEE}, \\H. Vincent Poor$^{2}$, \textit{Fellow,
IEEE}, and Hisashi Kobayashi$^{2}$, \textit{Life Fellow, IEEE}}

\maketitle

\footnotetext[0]{This research is supported in part by the
National Science Foundation under grants ANI-03-38807,
CCF-0440443, CCF-0448012, CNS-0417607, CNS-0427677, CNS-0417603,
and CCR-0440443, and in part by the New Jersey Center for Wireless
Telecommunications.} \footnotetext[1]{Mitsubishi Electric Research
Laboratories, 201 Broadway, Cambridge, MA 02139, USA, Tel:
(617)621-7500, Fax: (617)621-7550, email: gezici@merl.com}
\footnotetext[2]{Department of Electrical Engineering, Princeton
University, Princeton, NJ 08544, USA}

\maketitle

\begin{abstract}
The problem of choosing the optimal multipath components to be
employed at a minimum mean square error (MMSE) selective Rake
receiver is considered for an impulse radio ultra-wideband system.
First, the optimal finger selection problem is formulated as an
integer programming problem with a non-convex objective function.
Then, the objective function is approximated by a convex function
and the integer programming problem is solved by means of
constraint relaxation techniques. The proposed algorithms are
suboptimal due to the approximate objective function and the
constraint relaxation steps. However, they perform better than the
conventional finger selection algorithm, which is suboptimal since
it ignores the correlation between multipath components, and they
can get quite close to the optimal scheme that cannot be
implemented in practice due to its complexity. In addition to the
convex relaxation techniques, a genetic algorithm (GA) based
approach is proposed, which does not need any approximations or
integer relaxations. This iterative algorithm is based on the
direct evaluation of the objective function, and can achieve
near-optimal performance with a reasonable number of iterations.
Simulation results are presented to compare the performance of the
proposed finger selection algorithms with that of the conventional
and the optimal schemes.

\end{abstract}

\newpage

\section{Introduction}

Since the US Federal Communications Commission (FCC) approved the
limited use of ultra-wideband (UWB) technology \cite{FCC},
communications systems that employ UWB signals have drawn
considerable attention. A UWB signal is defined to be one that
possesses an absolute bandwidth larger than $500$MHz or a relative
bandwidth larger than 20\% and can coexist with incumbent systems
in the same frequency range due to its large spreading factor and
low power spectral density. UWB technology holds great promise for
a variety of applications such as short-range high-speed data
transmission and precise location estimation.

Commonly, impulse radio (IR) systems, which transmit very short
pulses with a low duty cycle, are employed to implement UWB
systems (\cite{scholtz}-\cite{andy}). In an IR system, a train of
pulses is sent and information is usually conveyed by the
positions or the amplitudes of the pulses, which correspond to
pulse position modulation (PPM) and pulse amplitude modulation
(PAM), respectively. In order to prevent catastrophic collisions
among different users and thus provide robustness against
multiple-access interference, each information symbol is
represented by a sequence of pulses, and the positions of the
pulses within that sequence are determined by a pseudo-random
time-hopping (TH) sequence specific to each user \cite{scholtz}.
The number $N_f$ of pulses representing one information symbol can
also be interpreted as pulse combining gain.

Typically, users in an IR-UWB system employ Rake receivers to
collect energy from different multipath components. A Rake
receiver combining all the paths of the incoming signal is called
an \textit{all-Rake} (\textit{ARake}) receiver. Since a UWB signal
has a very wide bandwidth, the number of resolvable multipath
components is usually very large. Hence, an ARake receiver is not
implemented in practice due to its complexity. However, it serves
as a benchmark for the performance of more practical Rake
receivers. A feasible implementation of multipath diversity
combining can be obtained by a \textit{selective-Rake}
(\textit{SRake}) receiver, which combines the $M$ best, out of
$L$, multipath components \cite{cassiICC02}. Those $M$ best
components are determined by a finger selection algorithm. For a
maximal ratio combining (MRC) Rake receiver, the paths with
highest signal-to-noise ratios (SNRs) are selected, which is an
optimal scheme in the absence of interfering users and
inter-symbol interference (ISI) \cite{1999_Win_Analysis HS-MRC
Rayleigh}-\cite{2000_Yue_GSC analysis}. For a minimum mean square
error (MMSE) Rake receiver, the ``conventional'' finger selection
algorithm can be defined as choosing the paths with highest
signal-to-interference-plus-noise ratios (SINRs). This
conventional scheme is not necessarily optimal since it ignores
the correlation of the noise terms at different multipath
components. The finger selection problem is also studied in the
context of CDMA downlink equalization, and recursive sequential
search (RSS) and heuristic arguments of interference cancellation
are proposed to determine finger locations
\cite{2003_Sui_CDMAfinger}. The RSS algorithm selects the fingers
one by one in a sequential manner, which reduces the
computationally complexity significantly; however, it may be quite
suboptimal depending on the correlation structure of the noise
components. The heuristic arguments are employed to determine
finger locations when the number of fingers is larger than the
number of multipath components, which is not applicable in our
case due to a large number of multipath components in a typical
UWB channel. In \cite{2005_Sui_CDMAfinger}, asymptotic optimality
of ``regular'' finger assignments is investigated. However, the
behavior of the algorithms for a small number of fingers, and
performance of other schemes without ``regular'' assignments are
not specified. Finally, the finger selection problem is considered
in \cite{2005_Zhiwei_MPfingerSel} for UWB systems, and a matching
pursuit-based technique with a quadratic constraint is proposed.
However, the performance of the algorithm depends heavily on a
parameter of the quadratic constraint, which needs to be
determined empirically.

In this paper, unlike the previous approaches, we provide a
complete optimization theoretical framework for the finger
selection problem for MMSE SRake receivers. First, we formulate
the optimal MMSE SRake as a nonconvex, integer-constrained
optimization, in which the aim is to choose the finger locations
of the receiver so as to maximize the overall SINR. While
computing the optimal finger selection is NP-hard, we present
several relaxation methods to turn the (approximate) problem into
convex optimization problems that can be very efficiently solved
by interior-point methods, which are polynomial-time in the worst
case, and are very fast in practice. These optimal finger
selection relaxations produce significantly higher average SINR
than the conventional one that ignores the correlations, and
represent a numerically efficient way to strike a balance between
SINR optimality and computational tractability. Moreover, we
propose a genetic algorithm (GA) based scheme, which performs
finger selection by iteratively evaluating the overall SINR
expression. Using this technique, near-optimal solutions can be
obtained in many cases with a degree of complexity that is much
lower than that of optimal search.

\newcounter{sec}
\setcounter{sec}{2} The remainder of this paper is organized as
follows. Section \Roman{sec} describes the transmitted and
received signal models in a multiuser frequency-selective
environment. The finger selection problem is formulated and the
optimal algorithm is described in Section
\setcounter{sec}{3}\Roman{sec}, followed by a brief description of
the conventional algorithm in Section
\setcounter{sec}{4}\Roman{sec}. In Section
\setcounter{sec}{5}\Roman{sec}, two convex relaxations of the
optimal finger selection algorithm, based on an approximate SINR
expression and integer constraint relaxation techniques, are
proposed. The GA-based approach is presented in Section
\setcounter{sec}{6}\Roman{sec} as an alternative to the suboptimal
schemes based on convex relaxations. Simulation results are
presented in Section \setcounter{sec}{7}\Roman{sec}, and
concluding remarks are made in the last section.

\section{Signal Model}

We consider a synchronous, binary phase shift keyed TH-IR system
with $K$ users, in which the transmitted signal from user $k$ is
represented by:
\begin{gather}\label{eq:tran1}
s^{(k)}_{{\rm{tx}}}(t)=\sqrt{\frac{E_k}{N_f}}\sum_{j=-\infty}^{\infty}d^{(k)}_j\,
b^{(k)}_{\lfloor j/N_f\rfloor}p_{{\rm{tx}}}(t-jT_f-c^{(k)}_jT_c),
\end{gather}
where $p_{{\rm{tx}}}(t)$ is the transmitted UWB pulse, $E_k$ is
the bit energy of user $k$, $T_f$ is the ``frame" time, $N_f$ is
the number of pulses representing one information symbol, and
$b^{(k)}_{\lfloor j/N_f \rfloor}\in \{+1,-1\}$ is the binary
information symbol transmitted by user $k$. In order to allow the
channel to be shared by many users and avoid catastrophic
collisions, a TH sequence $\{c^{(k)}_j\}$, where $c^{(k)}_j \in
\{0,1,...,N_c-1\}$, is assigned to each user. This TH sequence
provides an additional time shift of $c^{(k)}_jT_c$ seconds to the
$j$th pulse of the $k$th user where $T_c$ is the chip interval and
is chosen to satisfy $T_c\leq T_f/N_c$ in order to prevent the
pulses from overlapping. We assume $T_f=N_cT_c$ without loss of
generality. The random polarity codes $d^{(k)}_j$ are binary
random variables taking values $\pm1$ with equal probability
\cite{eran1}-\cite{paul}.

\begin{figure}
\begin{center}
\includegraphics[width=0.75\textwidth]{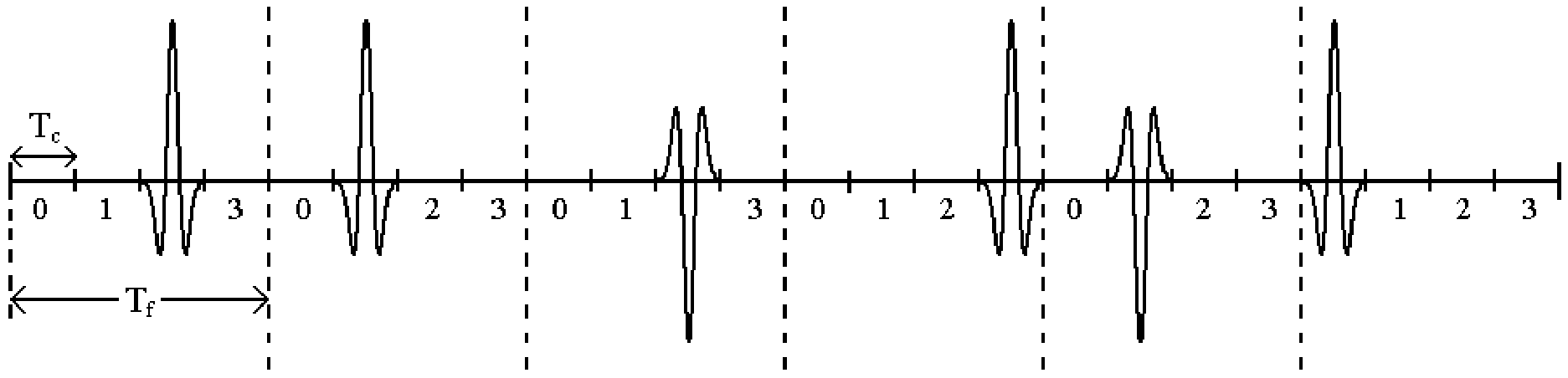}
\caption{An example time-hopping impulse radio signal with
pulse-based polarity randomization, where $N_f=6$, $N_c=4$, the
time hopping sequence is \{2,1,2,3,1,0\} and the polarity codes
are \{+1,+1,-1,+1,-1,+1\}.} \label{fig:TH_IR}
\end{center}
\end{figure}

An example TH-IR signal is shown in Figure \ref{fig:TH_IR}, where
six pulses are transmitted for each information symbol ($N_f=6$)
with the TH sequence $\{2,1,2,3,1,0\}$.

Consider the discrete presentation of the channel,
$\boldsymbol{\alpha}^{(k)}=[\alpha_1^{(k)}\cdots \alpha_L^{(k)}]$
for user $k$, where $L$ is assumed to be the number of multipath
components for each user, and $T_c$ is the multipath resolution.
Note that this channel model is quite general in that it can model
any channel of the form
$\sum_{l=1}^{\hat{L}}\hat{\alpha}_l^{(k)}\delta(t-\hat{\tau}_l^{(k)})$
if the channel is bandlimited to $1/T_c$ \cite{sinan_T-SP}. Then,
the received signal can be expressed as
\begin{gather}\label{eq:recSig}
r(t)=\sum_{k=1}^{K}\sqrt{\frac{E_k}{N_f}}\sum_{j=-\infty}^{\infty}\sum_{l=1}^{L}\alpha_l^{(k)}d^{(k)}_j\,
b^{(k)}_{\lfloor
j/N_f\rfloor}p_{{\rm{rx}}}(t-jT_f-c^{(k)}_jT_c-(l-1)T_c)+\sigma_nn(t),
\end{gather}
where $p_{{\rm{rx}}}(t)$ is the received unit-energy UWB pulse,
which is usually modelled as the derivative of $p_{{\rm{tx}}}(t)$
due to the effects of the receive antenna, and $n(t)$ is zero mean
white Gaussian noise with unit spectral density.

We assume that the TH sequence is constrained to the set
$\{0,1,\ldots,N_T-1\}$, where $N_T\leq N_c-L$, so that there is no
inter-frame interference (IFI). However, the proposed algorithms
are valid for scenarios with IFI as well, and this assumption is
made merely to simplify the expressions throughout the paper. From
the analysis in \cite{sinanUWBST04}, the results of this paper can
easily be extended to the IFI case as well.

\begin{figure}
\begin{center}
\includegraphics[width=0.75\textwidth]{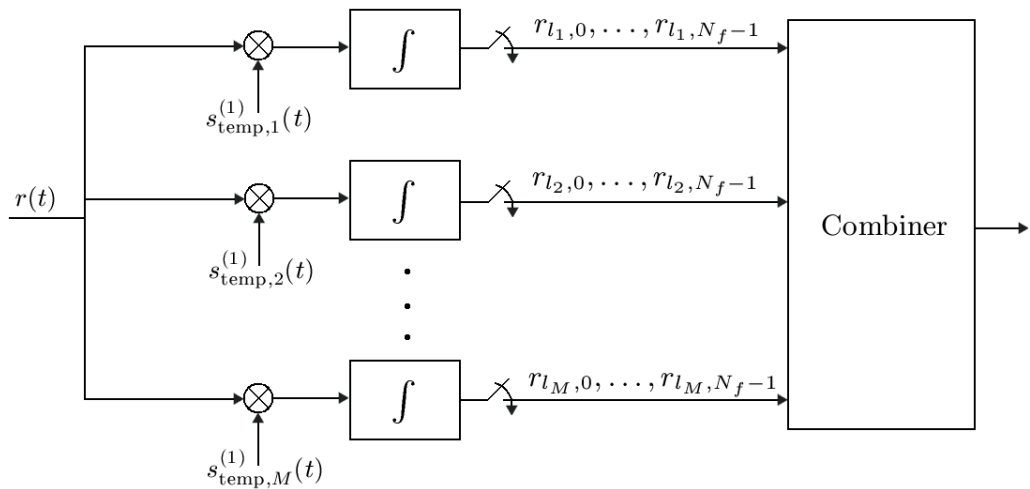}
\caption{The receiver structure. There are $M$ multipath
components, which are combined by the MMSE combiner.}
\label{fig:rec}
\end{center}
\end{figure}

Due to the high resolution of UWB signals, chip-rate and
frame-rate sampling are not very practical for such systems. In
order to have a lower sampling rate, the received signal can be
correlated with symbol-length template signals that enable
symbol-rate sampling of the correlator output
\cite{molischWPMC03}. The template signal for the $l$th path of
the incoming signal can be expressed as
\begin{gather}\label{eq:temp}
s_{{\rm{temp}},l}^{(1)}(t)=\sum_{j=iN_f}^{(i+1)N_f-1}d^{(1)}_j\,
p_{{\rm{rx}}}(t-jT_f-c^{(1)}_jT_c-(l-1)T_c),
\end{gather}
for the $i$th information symbol, where we consider user $1$ as
the desired user, without loss of generality. In other words, by
using a correlator for each multipath component that we want to
combine, we can have symbol-rate sampling at each branch, as shown
in Figure \ref{fig:rec}.

Note that the use of such template signals results in equal gain
combining (EGC) of different frame components. This may not be
optimal under some conditions (see \cite{sinanUWBST04} for
(sub)optimal schemes). However, it is very practical since it
facilitates symbol-rate sampling. Since we consider a system that
employs template signals of the form (\ref{eq:temp}), i.e. EGC of
frame components, it is sufficient to consider the problem of
selection of the optimal paths for just one frame. Hence, we
assume $N_f=1$ without loss of generality.

%FIG 2

Let $\mtL=\{l_1,\ldots,l_M\}$ denote the set of multipath
components that the receiver collects (Figure \ref{fig:rec}). At
each branch, the signal is effectively passed through a matched
filter (MF) matched to the related template signal in
(\ref{eq:temp}) and sampled once for each symbol. Then, the
discrete signal for the $l$th path can be expressed, for the $i$th
information symbol, as$^{2}$\footnotetext[2]{Note that the
dependence of $r_{l}$ on the index of the information symbol, $i$,
is not shown explicitly.}
\begin{gather}\label{eq:r_lj}
r_{l}=\ssT_{l}\aaa\bb_i+n_{l},
\end{gather}
for $l=l_1,\ldots,l_M$, where
$\aaa=\textrm{diag}\{\sqrt{E_1},\ldots,\sqrt{E_K}\}$,
$\bb_i=[b_i^{(1)}\cdots b_i^{(K)}]^T$ and
$n_{l}\sim\mathcal{N}(0\,,\,\sigma_n^2)$. $\ssV_{l}$ is a
$K\times1$ vector, which can be expressed as a sum of the desired
signal part (SP) and multiple-access interference (MAI) terms:
\begin{gather}\label{eq:sig_Vec}
\ssV_{l}=\ssV_{l}^{({\rm{SP}})}+\ssV_{l}^{({\rm{MAI}})},
\end{gather}
where the $k$th elements can be expressed as
\begin{align}\label{eq:sig_Vec_SP}
\left[\ssV_{l}^{({\rm{SP}})}\right]_k&=
\begin{cases}
\alpha_l^{(1)},\quad\quad k=1\\
0,\quad\quad\quad\,k=2,\ldots,K
\end{cases}\textrm{and}\\
\left[\ssV_{l}^{({\rm{MAI}})}\right]_k&=
\begin{cases}\label{eq:sig_Vec_MAI}
0,\quad\quad\quad\quad\quad\quad\quad\quad\quad\quad\,\,\, k=1\\
d_1^{(1)}d_1^{(k)}\sum_{m=1}^{L}\alpha_m^{(k)}I_{l,m}^{(k)},\quad
\quad k=2,\ldots,K
\end{cases},
\end{align}
with $I_{l,m}^{(k)}$ being the indicator function that is equal to
$1$ if the $m$th path of user $k$ collides with the $l$th path of
user $1$, and $0$ otherwise.

\section{Problem Formulation and Optimal Solution}

The problem is to choose the optimal set of multipath components,
$\mathcal{L}=\{l_1,\ldots,l_M\}$, that maximizes the overall SINR
of the system. In other words, we need to choose the best samples
from the $L$ received samples $r_l$, $l=1,\ldots,L$, as shown in
(\ref{eq:r_lj}).

To reformulate this combinatorial problem, we first define an
$M\times L$ selection matrix $\xxx$ as follows: $M$ of the columns
of $\xxx$ are the unit vectors $\ee_1,\ldots,\ee_M$ ($\ee_i$
having a $1$ at its $i$th position and zero elements for all other
entries), and the other columns are all zero vectors. The column
indices of the unit vectors determine the subset of the multipath
components that are selected. For example, for $L=4$ and $M=2$,
$\xxx=\begin{bmatrix}0\,\,1\,\,0\,\,0\\0\,\,0\,\,1\,\,0\end{bmatrix}$
chooses the second and third multipath components.

Using the selection matrix $\xxx$, we can express the vector of
received samples from $M$ multipath components as
\begin{gather}\label{eq:r1}
\rr=\xxx\sss\aaa\bb_i+\xxx\nn,
\end{gather}
where $\nn$ is the vector of thermal noise components
$\nn=[n_1\cdots n_L]^T$, and $\sss$ is the signature matrix given
by $\sss=[\ssV_1\cdots\ssV_{L}]^T$, with $\ssV_l$ as in
(\ref{eq:sig_Vec}).

From (\ref{eq:sig_Vec})-(\ref{eq:sig_Vec_MAI}), (\ref{eq:r1}) can
be expressed as
\begin{gather}
\rr=b_i^{(1)}\sqrt{E_1}\xxx\boldsymbol{\alpha}^{(1)}+\xxx\sss^{({\rm{MAI}})}\aaa\bb_i+\xxx\nn,
\end{gather}
where $\sss^{({\rm{MAI}})}$ is the MAI part of the signature
matrix $\sss$.

Then, the linear MMSE receiver can be expressed as
\begin{gather}
\hat{b}_i=\textrm{sign}\{\boldsymbol{\theta}^T\rr\},
\end{gather}
where the MMSE weight vector is given by \cite{verdu}
\begin{gather}
\boldsymbol{\theta}=\rrr^{-1}\xxx\boldsymbol{\alpha}^{(1)},
\end{gather}
with $\rrr$ being the correlation matrix of the noise term:
\begin{gather}
\rrr=\xxx\sss^{({\rm{MAI}})}\aaa^2(\sss^{({\rm{MAI}})})^T\xxx^T+\sigma_n^2\iii.
\end{gather}

The overall SINR of the system can be expressed as
\begin{gather}\label{eq:SINR}
{\rm{SINR}}(\xxx)=\frac{E_1}{\sigma_n^2}(\boldsymbol{\alpha}^{(1)})^T\xxx^T
\left(\iii+\frac{1}{\sigma_n^2}\xxx\sss^{({\rm{MAI}})}\aaa^2(\sss^{({\rm{MAI}})})^T\xxx^T\right)^{-1}\xxx\boldsymbol{\alpha}^{(1)}.
\end{gather}
Hence, the optimal finger selection problem can be formulated as
finding $\xxx$ that solves the following problem:
\begin{gather}\label{eq:optProb}
\textrm{maximize}\,\,{\rm{SINR}}(\xxx),
\end{gather}
subject to the constraint that $\xxx$ has the previously defined
structure. Note that the objective function to be maximized is not
concave and the optimization variable $\xxx$ takes binary values,
with the previously defined structure. In other words, two major
difficulties arise in solving (\ref{eq:optProb}) globally:
nonconvex optimization and integer constraints. Either makes the
problem NP-hard. Therefore, it is an intractable optimization
problem in this general form.

\section{Conventional Algorithm}

Instead of the solving the problem in (\ref{eq:optProb}), the
``conventional" finger selection algorithm chooses the $M$ paths
with largest individual SINRs, where the SINR for the $l$th path
can be expressed as
\begin{gather}\label{eq:indivSINR}
{\rm{SINR}}_l=\frac{E_1(\alpha_l^{(1)})^2}{(\ssV_l^{({\rm{MAI}})})^T\aaa^2\ssV_l^{({\rm{MAI}})}+\sigma_n^2},
\end{gather}
for $l=1,\ldots,L$.

This algorithm is not optimal because it ignores the correlation
of the noise components of different paths. Therefore, it does not
always maximize the overall SINR of the system given in
(\ref{eq:SINR}). For example, the contribution of two highly
correlated strong paths to the overall SINR might be worse than
the contribution of one strong and one relatively weaker, but
uncorrelated, paths. The correlation between the multipath
components is the result of the MAI from the interfering users in
the system.

\section{Relaxations of Optimal Finger Selection}\label{sec:Subopt}

Since the optimal solution in (\ref{eq:optProb}) is quite
difficult, we first consider an approximation of the objective
function in (\ref{eq:SINR}). When the eigenvalues of
$\frac{1}{\sigma_n^2}\xxx\sss^{({\rm{MAI}})}\aaa^2(\sss^{({\rm{MAI}})})^T\xxx^T$
are considerably smaller than $1$, which occurs when the MAI is
not very strong compared to the thermal noise, we can approximate
the SINR expression in (\ref{eq:SINR}) as
follows$^{3}$\footnotetext[3]{More accurate approximations can be
obtained by using higher order series expansions for the matrix
inverse in (\ref{eq:SINR}). However, the solution of the
optimization problem does not lend itself to low complexity
solutions in those cases.}:
\begin{gather}\label{eq:SINR2}
{\rm{SINR}}(\xxx)\approx\frac{E_1}{\sigma_n^2}(\boldsymbol{\alpha}^{(1)})^T\xxx^T
\left(\iii-\frac{1}{\sigma_n^2}\xxx\sss^{({\rm{MAI}})}\aaa^2(\sss^{({\rm{MAI}})})^T\xxx^T\right)\xxx\boldsymbol{\alpha}^{(1)},
\end{gather}
which can be expressed as
\begin{gather}\label{eq:SINR3}
{\rm{SINR}}(\xxx)\approx\frac{E_1}{\sigma_n^2}\{\,(\boldsymbol{\alpha}^{(1)})^T\xxx^T\xxx\boldsymbol{\alpha}^{(1)}
-\frac{1}{\sigma_n^2}\boldsymbol{\alpha}^{(1)}\xxx^T\xxx\sss^{({\rm{MAI}})}\aaa^2(\sss^{({\rm{MAI}})})^T\xxx^T\xxx\boldsymbol{\alpha}^{(1)}\,\}.
\end{gather}
Note that the approximate SINR expression depends on $\xxx$ only
through $\xxx^T\xxx$. Defining $\xx=[x_1\cdots x_L]^T$ as the
diagonal elements of $\xxx^T\xxx$,
$\xx=\textrm{diag}\{\xxx^T\xxx\}$, we have $x_i=1$ if the $i$th
path is selected, and $x_i=0$ otherwise; and
$\sum_{i=1}^{L}x_i=M$. Then, we obtain, after some manipulation,
\begin{gather}\label{eq:SINR4}
{\rm{SINR}}(\xx)=\frac{E_1}{\sigma_n^2}\left\{\qq^T\xx-\frac{1}{\sigma_n^2}\xx^T\ppp\xx\right\},
\end{gather}
where $\qq=[(\alpha_1^{(1)})^2\cdots(\alpha_L^{(1)})^2]^T$ and
$\ppp=\textrm{diag}\{\alpha_1^{(1)}\cdots\alpha_L^{(1)}\}\sss^{({\rm{MAI}})}\aaa^2(\sss^{({\rm{MAI}})})^T\textrm{diag}\{\alpha_1^{(1)}\cdots\alpha_L^{(1)}\}$.

Then, we can formulate the finger selection problem as follows:
\begin{align}\label{eq:subopt}\nonumber
\textrm{minimize}\,\quad&\frac{1}{\sigma_n^2}\xx^T\ppp\xx-\xx^T\qq\\\nonumber
\textrm{subject to}\quad&\xx^T\one=M,\\
&x_i\in\{0,1\},\quad i=1,\ldots,L.
\end{align}

Note that the objective function is convex since $\ppp$ is
positive definite, and that the first constraint is linear.
However, the integer constraint increases the complexity of the
problem. The common way to approximate the solution of an integer
constraint problem is to use \textit{constraint relaxation}. Then,
the optimizer will be a continuous value instead of being binary
and the problem (\ref{eq:subopt}) will be convex.
%Therefore, the solution of the finger selection problem
%will be to choose the paths corresponding to $M$ largest elements
%of the optimizer.
Over the past decade, both powerful theory and efficient numerical
algorithms have been developed for nonlinear convex optimization.
It is now recognized that the watershed between ``easy'' and
``difficult'' optimization problems is not linearity but
convexity. For example, the interior-point algorithms for
nonlinear convex optimization are very fast, both in theory and in
practice: they are provably polynomial time in theory, and in
practice usually take about $25$-$40$ times the computational load
of solving a least-squares problem of the same dimension to reach
a global optimum of a constrained convex optimization problem
\cite{optTextbook}, \cite{Nesterov_IntPointBOOK}. The
interior-point methods solve convex optimization problems with
inequality constraints by applying Newton's method to a sequence
of equality constrained problems, where the Newton's method is a
kind of descent algorithm with the descent direction given by the
Newton step \cite{optTextbook}.

We consider two different relaxation techniques in the following
subsections.

\subsection{Case-1: Relaxation to Sphere}\label{sec:sphere}

Consider the relaxation of the integer constraint in
(\ref{eq:subopt}) to a sphere that passes through all possible
integer values. Then, the relaxed problem becomes
\begin{align}\label{eq:subopt2}\nonumber
\textrm{minimize}\,\quad&\frac{1}{\sigma_n^2}\xx^T\ppp\xx-\xx^T\qq\\\nonumber
\textrm{subject to}\quad&\,\xx^T\one=M,\\
&(2\xx-\one)^T(2\xx-\one)\leq L.
\end{align}
Note that the problem becomes a convex quadratically constrained
quadratic programming (QCQP) problem \cite{optTextbook}. Hence it
can be solved for global optimality using interior-point
algorithms in polynomial time.

\subsection{Case-2: Relaxation to Hypercube}\label{sec:hypercube}

As an alternative approach, we can relax the integer constraint in
(\ref{eq:subopt}) to a hypercube constraint and obtain
\begin{align}\label{eq:subopt3}\nonumber
\textrm{minimize}\,\quad&\frac{1}{\sigma_n^2}\xx^T\ppp\xx-\xx^T\qq\\\nonumber
\textrm{subject to}\quad&\,\xx^T\one=M,\\
&\,\xx\in[0,1]^L,
\end{align}
where the hypercube constraint can be expressed as
$\xx\succeq\bold{0}$ and $\xx\preceq\one$, with $\yy\succeq\zz$
meaning that $y_1\geq z_1,\ldots,y_L\geq z_L$. Note that the
problem is now a linearly constrained quadratic programming (LCQP)
problem, and can be solved by interior-point algorithms
\cite{optTextbook} for the optimizer $\xx^{*}$.

\subsection{Dual Methods}

We can also consider the dual problems. For the relaxation to the
sphere considered in Section \ref{sec:sphere}, the Lagrangian for
(\ref{eq:subopt2}) can be obtained as
\begin{gather}
\mathcal{L}(\xx,\lambda,\nu)=\xx^T\left(\frac{1}{\sigma_n^2}\ppp+\nu\iii\right)\xx-\xx^T(\qq-\lambda\one+\nu\one)-M\lambda,
\end{gather}
where $\lambda\in\mathcal{R}$ and $\nu\in\mathcal{R}^{+}$.

After some manipulation, the Lagrange dual function, which is the
Lagrangian maximized over the primal variable $\xx$, can be
expressed as
\begin{gather}
g(\lambda,\nu)=-\frac{1}{4}\,[\qq+(\nu-\lambda)\one]^T\left(\frac{1}{\sigma_n^2}\ppp+\nu\iii\right)^{-1}
[\qq+(\nu-\lambda)\one]-M\lambda,
\end{gather}
Then, the dual problem becomes
\begin{align}\label{eq:subopt2_dual}
&\textrm{minimize}\quad\frac{1}{4}[\qq+(\nu-\lambda)\one]^T\left(\frac{1}{\sigma_n^2}\ppp+\nu\iii\right)^{-1}[\qq+(\nu-\lambda)\one]+M\lambda\\
&\textrm{subject to}\quad\nu\geq 0,
\end{align}
which can be solved for optimal $\lambda$ and $\nu$ by interior
point methods. Or, more simply, the unconstrained problem
(\ref{eq:subopt2_dual}) can be solved using a gradient descent
algorithm, and then the optimizer $\bar{\nu}$ is mapped to
$\nu^{*}=\max\{0,\bar{\nu}\}$.

After solving for optimal $\lambda$ and $\mu$, the optimizer
$\xx^{*}$ is obtained as
\begin{gather}
\xx^{*}=\frac{1}{2}\left(\frac{1}{\sigma_n^2}\ppp+\nu^{*}\iii\right)^{-1}[\qq+(\nu^{*}-\lambda^{*})\one].
\end{gather}

Note that the dual problem (\ref{eq:subopt2_dual}) has two
variables, $\lambda$ and $\nu$, to optimize, compared to $L$
variables, the components of $\xx$, in the primal problem
(\ref{eq:subopt2}). However, an $L\times L$ matrix must be
inverted for each iteration of the optimization of
(\ref{eq:subopt2_dual}). Therefore, the primal problem may be
preferred over the dual problem in certain cases.

Similarly, the dual problem for the relaxation in Section
\ref{sec:hypercube} can be obtained from (\ref{eq:subopt3}) as
\begin{align}\label{eq:subopt3_dual}
&\textrm{minimize}\quad\frac{\sigma_n^2}{4}(\qq+\MU-\NU-\lambda\one)^T\ppp^{-1}(\qq+\MU-\NU-\lambda\one)+M\lambda+\NU^T\one\\
&\textrm{subject to}\quad\MU,\NU\succeq \zero.
\end{align}

It is observed from (\ref{eq:subopt3_dual}) that there are $2L+1$
variables and also $L\times L$ matrix inversion operations for the
solution of the dual problem. Therefore, the simpler primal
problem (\ref{eq:subopt3}) is considered in the simulations.

\subsection{Selection of Finger Locations}

After solving the approximate problem (\ref{eq:subopt}) by means
of integer relaxation techniques mentioned above, the finger
location estimates are obtained by the indices of the $M$ largest
elements of the optimizer $\xx^{*}$.

Both the approximation of the SINR expression by (\ref{eq:SINR2})
and the integer relaxation steps result in the suboptimality of
the solution. Therefore, it may not be very close to the optimal
solution in some cases. However, it is expected to perform better
than the conventional algorithm most of the time, since it
considers the correlation between the multipath components.
However, it is not guaranteed that the algorithms based on the
convex relaxations of optimal finger selection always beat the
conventional one. Since the conventional algorithm is very easy to
implement, we can consider a hybrid algorithm in which the final
estimate of the convex relaxation algorithm is compared with that
of the conventional one, and the one that minimizes the exact SINR
expression in (\ref{eq:SINR}) is chosen as the final estimate. In
this way, the resulting hybrid suboptimal algorithm can get closer
to the optimal solution.

\section{Finger Selection Using Genetic Algorithms}

The algorithms in the previous section convert the optimal finger
selection problem into a convex problem by approximating the SINR
expression in (\ref{eq:SINR}) by a concave function in
(\ref{eq:SINR4}), and by employing the relaxation techniques on
the integer constraints. As another way to solve the finger
selection problem, we propose a GA based approach, which directly
uses the exact SINR expression in (\ref{eq:SINR}), and does not
employ any relaxation techniques.

\subsection{Genetic Algorithm}

The GA is an iterative technique for searching for the global
optimum of a cost function \cite{GA1}. The name comes from the
fact that the algorithm models the natural selection and survival
of the fittest \cite{GA2}.

The GA starts with a population of chromosomes, where each
chromosome is represented by a binary
string$^{5}$\footnotetext[5]{Although we consider only the binary
GA, continuous parameter GAs are also available \cite{GA1}.}. Let
$N_{{\rm{ipop}}}$ denote the number of chromosomes in this
population. Then, the fittest $N_{{\rm{pop}}}$ of these
chromosomes are selected, according to a fitness function. After
that, the fittest $N_{{\rm{good}}}$ chromosomes, which are also
called the ``parents'', are selected and paired among themselves
(\textit{pairing} step). From each chromosome pair, two new
chromosomes are generated, which is called the \textit{mating}
step. In other words, the new population consists of
$N_{{\rm{good}}}$ parent chromosomes and $N_{{\rm{good}}}$
children generated from the parents by mating. After the mating
step, the \textit{mutation} stage follows, where some chromosomes
(the fittest one in the population can be excluded) are chosen
randomly and are slightly modified; that is, some bits in the
selected binary string are flipped. After that, the pairing,
mating and mutation steps are repeated until a threshold criterion
is met.

The GA has been applied to a variety of problems in different
areas \cite{GA1}-\cite{GA3}. Also, it has recently been employed
in the multiuser detection problem
\cite{Juntti_1997}-\cite{Hanzo_2004}. The main characteristics of
the GA algorithm is that it can get close to the optimal solution
with low complexity, if the steps of the algorithm are designed
appropriately.

\subsection{Finger Selection via the GA}

In order to be able to employ the GA for the finger selection
problem we need to consider how to represent the chromosomes, and
how to implement the steps of the iterative optimization scheme.

A natural way to represent a chromosome is to consider the
``assignment vector'' $\xx$ in (\ref{eq:SINR4}), which denotes the
assignments of the multipath components to the $M$ fingers of the
RAKE receiver. In other words, $x_i=1$ if the $i$th path is
selected, and $x_i=0$ otherwise; and $\sum_{i=1}^{L}x_i=M$.

Also, the fitness function that should be maximized can be the
SINR expression given by (\ref{eq:SINR}). Note that, given a value
of $\xx$, ${\rm{SINR}}(\xxx)$ can be uniquely evaluated. By
choosing this fitness function, the fittest chromosomes of the
population correspond to the assignment vectors with the largest
SINR values.

Now the pairing, mating and mutation steps need to be designed for
the finger selection problem:

\subsubsection{Pairing} The assignments to be paired among themselves
are chosen according to a weighted random pairing scheme
\cite{GA1}, where each assignment is chosen with a probability
that is proportional to its SINR value. In this way, the
assignments with large SINR values have a greater chance of being
chosen as the parents for the new assignments.

\subsubsection{Mating} From each assignment pair, the two new
pairs are generated in the following manner: Let $\xx_1$ and $\xx_2$
denote two finger assignments, and let $\pp_{\xx_1}$ and
$\pp_{\xx_2}$ consist of the indices of the multipath components
chosen as the RAKE fingers. Then, the indices of the new assignments
are chosen randomly from the vector
$\pp=[\pp_{\xx_1}\,\,\pp_{\xx_2}]$. If the new assignment is the
same as $\xx_1$ or $\xx_2$, the procedure is repeated for that
assignment. For example, consider a case where $L=10$ and $M=4$. If
$\xx_1=[1\,\,0\,\,0\,\,1\,\,0\,\,0\,\,1\,\,1\,\,0\,\,0]$
($\pp_{\xx_1}=[1\,\,4\,\,7\,\,8]$) and
$\xx_2=[0\,\,1\,\,0\,\,1\,\,0\,\,1\,\,0\,\,0\,\,1\,\,0]$
($\pp_{\xx_2}=[2\,\,4\,\,6\,\,9]$), then the new assignments are
chosen randomly from the set
$\pp=[1\,\,4\,\,7\,\,8\,\,2\,\,4\,\,6\,\,9]$. For example, the new
assignments (children) could be
$\xx_3=[1\,\,1\,\,0\,\,1\,\,0\,\,0\,\,0\,\,0\,\,1\,\,0]$
($\pp_{\xx_3}=[1\,\,2\,\,4\,\,9]$) and
$\xx_4=[0\,\,0\,\,0\,\,1\,\,0\,\,1\,\,1\,\,0\,\,1\,\,0]$
($\pp_{\xx_4}=[4\,\,6\,\,7\,\,9]$).

Note that by designing such a mating algorithm, we make sure that
a multipath component that is selected by both parents has a
larger probability of being selected by the new assignment than a
multipath component that is selected by only one parent does.

\subsubsection{Mutation} In the mutation step, an
assignment, except the best one (the one with the highest SINR),
is randomly selected, and one $1$ and one $0$ of that assignment
are randomly chosen and flipped. This mutation operation can be
repeated a number of times for each iteration. The number of
mutations can be determined beforehand, or it might be defined as
a random variable.

Now, we can summarize our GA-based finger selection scheme as
follows:
\begin{itemize}
\item Generate $N_{{\rm{ipop}}}$ different assignments randomly.
\item Select $N_{{\rm{pop}}}$ of them with the largest SINR
values. \item \textbf{Pairing:} Pair $N_{{\rm{good}}}$ of the
finger assignments according to the weighted random scheme. \item
\textbf{Mating:} Generate two new assignments from each pair.
\item \textbf{Mutation:} Change the finger locations of some
assignments randomly except for the best assignment. \item Choose
the assignment with the highest SINR if the threshold criterion is
met; go to the pairing step otherwise.
\end{itemize}

In the simulations, we stop the algorithm after a certain number
of iterations. In other words, the threshold criterion is that the
number of iterations exceeds a given value. As the number of
iterations increases, the performance of the algorithm increases,
as well. The other parameters that determine the tradeoff between
complexity and performance are $N_{{\rm{ipop}}}$,
$N_{{\rm{pop}}}$, $N_{{\rm{good}}}$, and the number of mutations
at each iteration.

\section{Simulation Results}

Simulations have been performed to evaluate the performance of
various finger selection algorithms for an IR-UWB system with
$N_c=20$ and $N_f=1$. In these simulations, there are five equal
energy users in the system ($K=5$) and the users' TH and polarity
codes are randomly generated. We model the channel coefficients as
$\alpha_l=\textrm{sign}(\alpha_l)|\alpha_l|$ for $l=1,\ldots,L$,
where $\textrm{sign}(\alpha_l)$ is $\pm1$ with equal probability
and $|\alpha_l|$ is distributed lognormally as
$\mathcal{LN}(\mu_l,\sigma^2)$. Also the energy of the taps is
exponentially decaying as
$\textrm{E}\{|\alpha_l|^2\}=\Omega_0e^{-\lambda(l-1)}$, where
$\lambda$ is the decay factor and
$\sum_{l=1}^{L}\textrm{E}\{|\alpha_l|^2\}=1$ (so
$\Omega_0=(1-e^{-\lambda})/(1-e^{-\lambda L})$). For the channel
parameters, we choose $\lambda=0.1$, $\sigma^2=0.5$ and $\mu_l$
can be calculated from
$\mu_l=0.5\left[\textrm{ln}(\frac{1-e^{-\lambda}}{1-e^{-\lambda
L}})-\lambda(l-1)-2\sigma^2\right]$, for $l=1,\ldots,L$. We
average the overall SINR of the system over different realizations
of channel coefficients, TH and polarity codes of the users.

% FIG 3

\begin{figure}
\begin{center}
\includegraphics[width=0.95\textwidth]{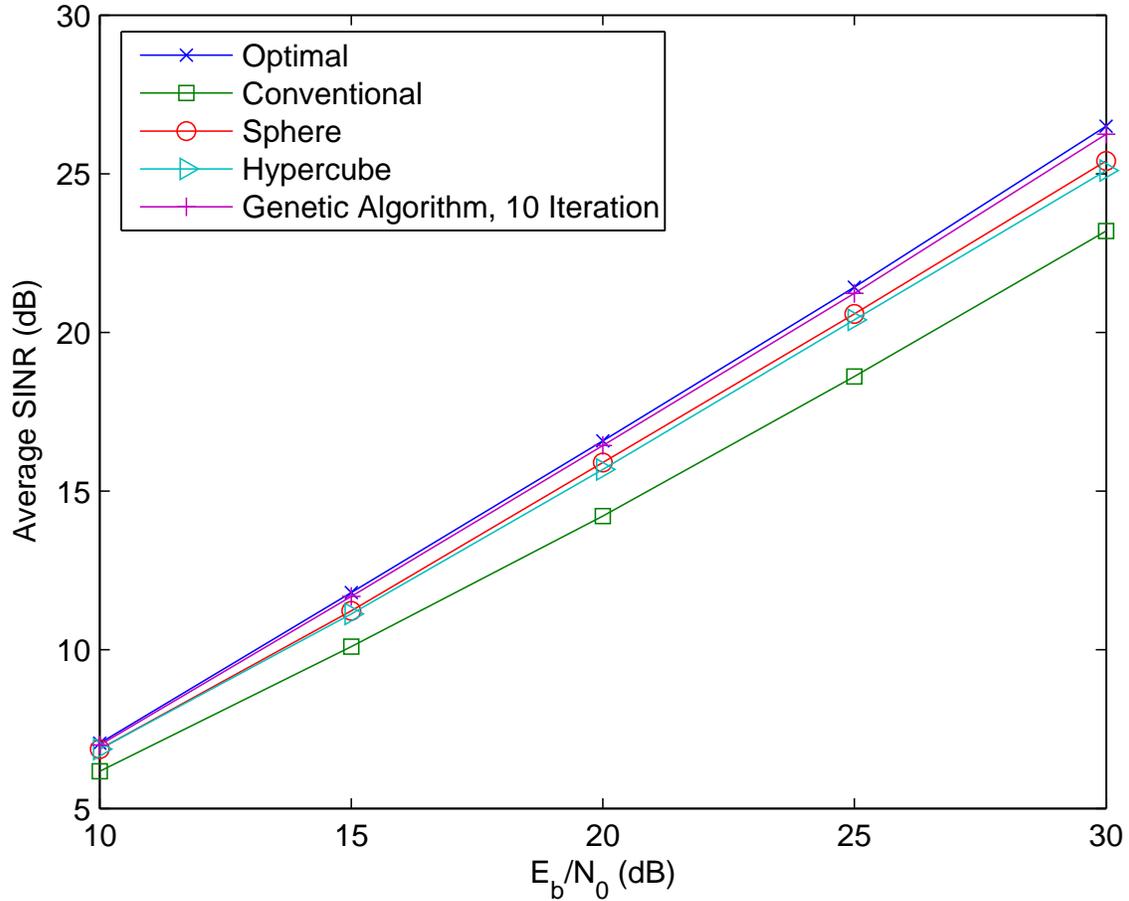}
\caption{\small Average $\rm{SINR}$ versus $E_b/N_0$ for $M=5$
fingers, where $E_b$ is the bit energy. The channel has $L=15$
multipath components and the taps are exponentially decaying. The
IR-UWB system has $N_c=20$ chips per frame and $N_f=1$ frame per
symbol. There are $5$ equal energy users in the system and random
TH and polarity codes are used.} \label{fig:SINR_vs_EbNo}
\end{center}
\end{figure}

In Figure \ref{fig:SINR_vs_EbNo}, we plot the average SINR of the
system for different noise variances when $M=5$ fingers are to be
chosen out of $L=15$ multipath components. For the GA,
$N_{{\rm{ipop}}}=32$, $N_{{\rm{pop}}}=16$, and $N_{{\rm{good}}}=8$
are used, and $8$ mutations are performed at each iteration. As is
observed from the figure, the convex relaxations of optimal finger
selection and the GA based scheme perform considerably better than
the conventional scheme, and the GA get very close to the optimal
exhaustive search scheme after $10$ iterations. Note that the gain
achieved by using the proposed algorithms over the conventional
one increases as the thermal noise decreases. This is because when
the thermal noise becomes less significant, the MAI becomes
dominant, and the conventional technique gets worse since it
ignores the correlation between the MAI noise terms when choosing
the fingers.

% FIG 4

\begin{figure}
\begin{center}
\includegraphics[width=0.95\textwidth]{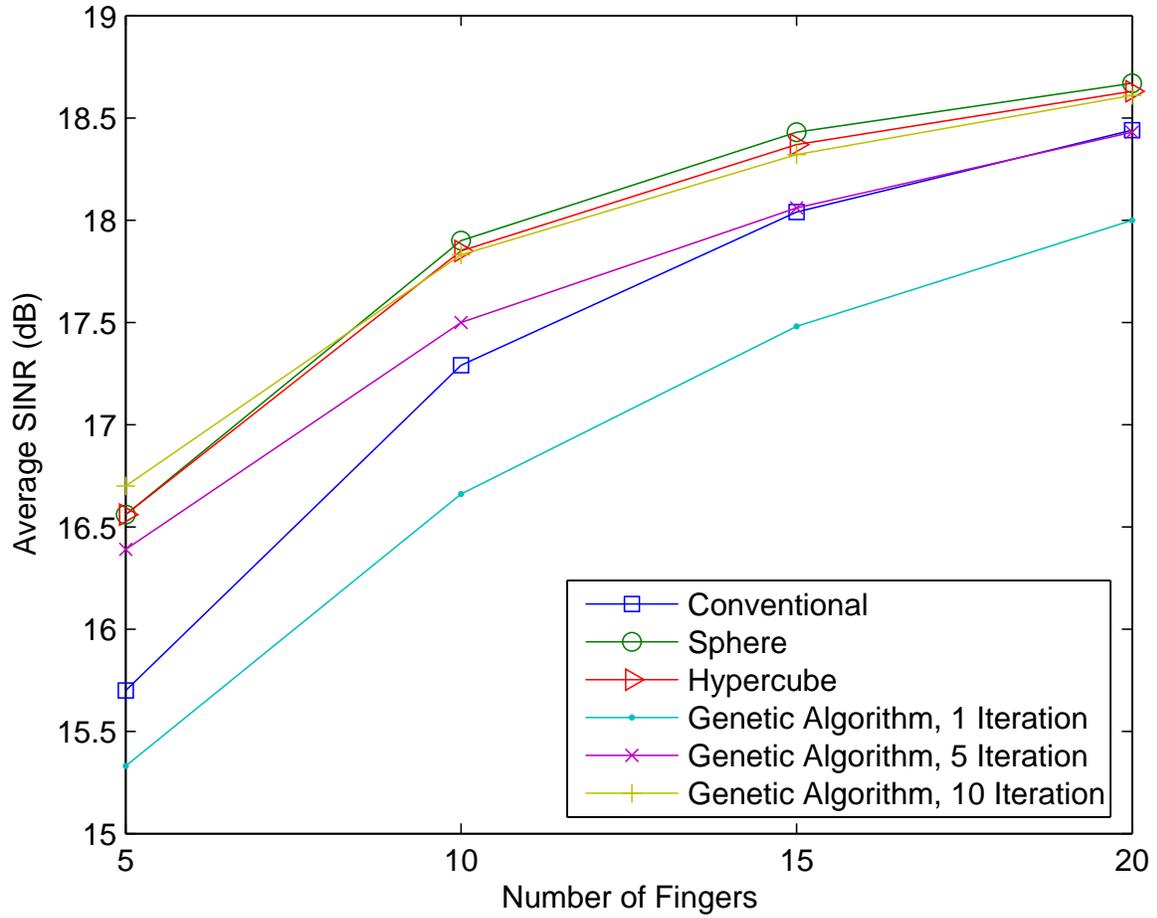}
\caption{\small Average $\rm{SINR}$ versus number of fingers $M$,
for $E_b/N_0=20$dB, $N_c=75$ and $L=50$. All the other parameters
are the same as those for Figure \ref{fig:SINR_vs_EbNo}.}
\label{fig:SINR_vs_M_2}
\end{center}
\end{figure}

Next, we plot the SINR of the proposed suboptimal and conventional
techniques for different numbers of fingers in Figure
\ref{fig:SINR_vs_M_2}, where there are $50$ multipath components
and $E_b/N_0=20$. The number of chips per frame, $N_c$, is set to
$75$, and all other parameters are kept the same as before. In
this case, the optimal algorithm takes a very long time to
simulate since it needs to perform exhaustive search over many
different finger combinations and therefore it was not
implemented. The improvement using convex relaxations of optimal
finger selection over the conventional technique decreases as $M$
increases since the channel is exponentially decaying and most of
the significant multipath components are already combined by all
the algorithms. Also, the GA based scheme performs very close to
the suboptimal schemes using convex relaxations after $10$
iterations with $N_{{\rm{ipop}}}=128$, $N_{{\rm{pop}}}=64$,
$N_{{\rm{good}}}=32$, and $32$ mutations.

% FIG 5

\begin{figure}
\begin{center}
\includegraphics[width=0.95\textwidth]{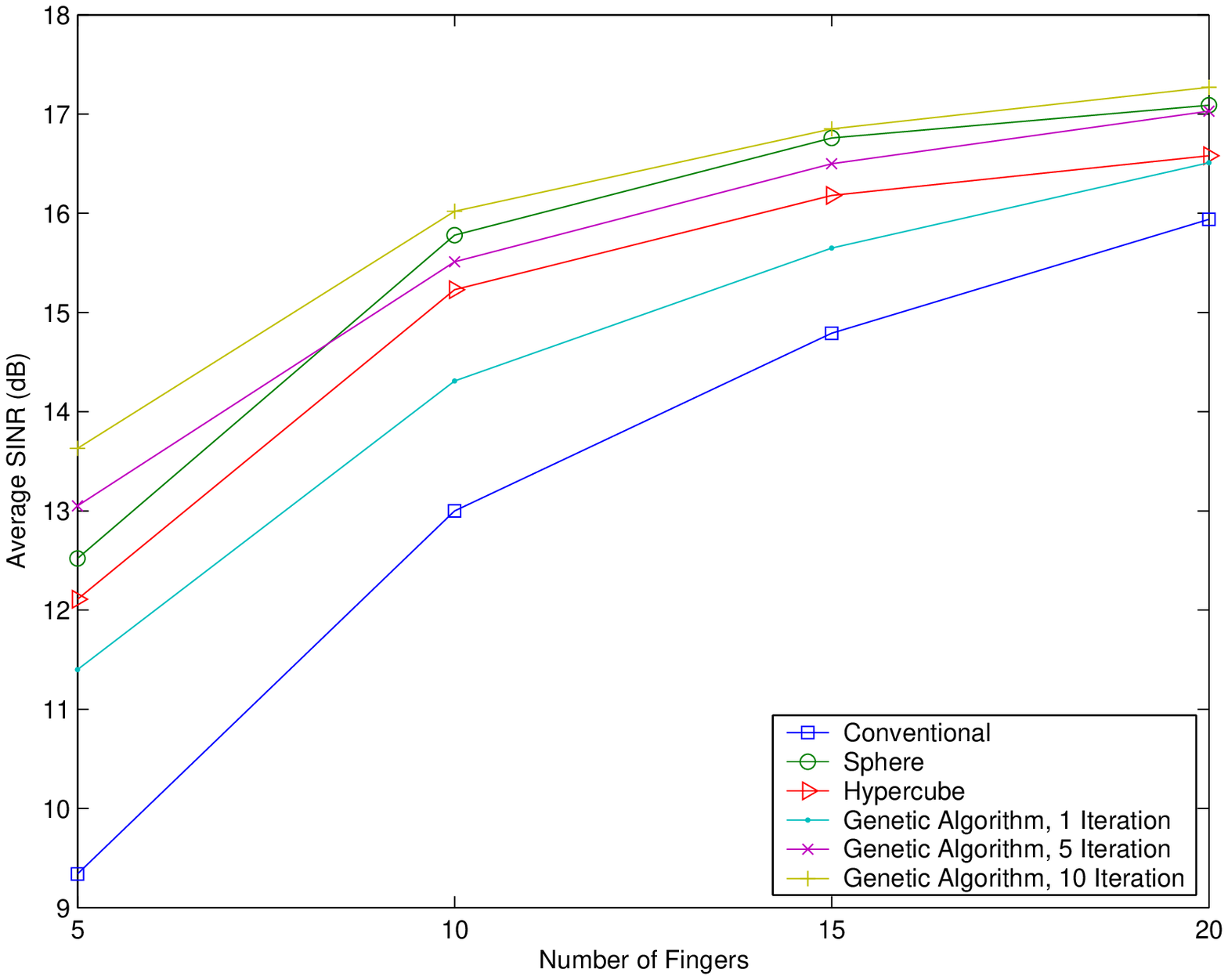}
\caption{\small Average $\rm{SINR}$ versus number of fingers $M$.
There are $10$ users with each interferer having $10$dB more power
than the desired user. All the other parameters are the same as
those for Figure \ref{fig:SINR_vs_M_2}.}
\label{fig:SINR_vs_M_MAIlim}
\end{center}
\end{figure}

Finally, we consider an MAI-limited scenario, in which there are
$10$ users with $E_1=1$ and $E_k=10$ $\forall k\ne1$, and all the
parameters are as in the previous case. Then, as shown in Figure
\ref{fig:SINR_vs_M_MAIlim}, the improvement by using the
suboptimal finger selection algorithms increase significantly. The
main reason for this is that the suboptimal algorithms consider
(approximately) the correlation caused by MAI whereas the
conventional scheme simply ignores it.

\section{Concluding Remarks}

Optimal and suboptimal finger selection algorithms for MMSE-SRake
receivers in an IR-UWB system have been considered. Since UWB
systems have large numbers of multipath components, only a subset
of those components can be used due to complexity constraints.
Therefore, the selection of the optimal subset of multipath
components is important for the performance of the receiver. We
have shown that the optimal solution to this finger selection
problem requires exhaustive search which becomes prohibitive for
UWB systems. Therefore, we have proposed approximate solutions of
the problem based on Taylor series approximation and integer
constraint relaxations. Using two different integer relaxation
approaches, we have introduced two convex relaxations of the
optimal finger selection algorithm.

Moreover, we have proposed a GA based iterative finger selection
scheme, which depends on the direct evaluation of the objective
function. In each iteration, the set of possible finger
assignments is updated in search of the best assignment according
to the proposed GA stages.

The three contributions of this paper are the formulation of the
optimization problem, the convex relaxations, and the GA based
scheme. In the first, the formulation is globally optimal, but the
solution methods for this non-convex nonlinear integer constrained
optimization must use heuristics due to its prohibitive
computational complexity. In the second, the suboptimal schemes
focus on relaxations where globally optimal solutions are
guaranteed, but the problem statement is relaxed. In the third,
the problem statement remains the same as the original,
intractable NP hard one, but the solution method is by local
search heuristics.

\newpage

\begin{thebibliography}{99}

\bibitem{FCC} U. S. Federal Communications Commission, FCC 02-48: First Report and Order.

\bibitem{scholtz} M. Z. Win and R. A. Scholtz, ``Impulse radio:
How it works," \emph{IEEE Communications Letters,} 2(2): pp.
36-38, Feb. 1998.

\bibitem{win} M. Z. Win and R. A. Scholtz, ``Ultra-wide bandwidth time-hopping spread-spectrum impulse
radio for wireless multiple-access communications," \emph{IEEE
Transactions on Communications}, vol. 48, pp. 679-691, April 2000.

\bibitem{ramirez} F. Ramirez Mireless, ``On the performance of ultra-wideband signals in Gaussian noise and dense
multipath,'' \emph{IEEE Transactions on Vehicular Technology,}
vol. 50, no. 1, pp. 244-249, Jan. 2001.

\bibitem{scholtz2} R. A. Scholtz, ``Multiple access with time-hopping impulse
modulation," \emph{Proc. IEEE Military Communications Conference
(MILCOM 1993),} vol. 2, pp. 447-450, Boston, MA, Oct. 1993.

\bibitem{andy} D. Cassioli, M. Z. Win and A. F. Molisch, ``The
ultra-wide bandwidth indoor channel: From statistical model to
simulations," \emph{IEEE Journal on Selected Areas in
Communications}, vol. 20, pp. 1247-1257, Aug. 2002.

\bibitem{cassiICC02} D. Cassioli, M. Z. Win, F. Vatalaro and A. F.
Molisch, ``Performance of low-complexity RAKE reception in a
realistic UWB channel," \emph{Proc. IEEE International Conference
on Communications (ICC 2002),} vol. 2, pp. 763-767, New York, NY,
April 28-May 2, 2002.




\bibitem{1999_Win_Analysis HS-MRC Rayleigh} M. Z. Win and J. H.
Winters, ``Analysis of hybrid selection/maximal-ratio combining of
diversity branches with unequal SNR in Rayleigh fading,''
\emph{Proc. IEEE 49th Vehicular Technology Conference (VTC
1999-Spring),} vol. 1, pp. 215-220, Houston, TX, May 16-20, 1999.


\bibitem{1998_Milstein_Avg SNR GSC} N. Kong and L. B. Milstein,
``Combined average SNR of a generalized diversity selection
combining scheme,'' \emph{Proc. IEEE International Conference on
Communications (ICC 1998),} vol. 3, pp. 1556-1560, Atlanta, GA,
June 7-11, 1998.

\bibitem{2000_Yue_GSC analysis} L. Yue, ``Analysis of generalized selection combining
techniques,'' \emph{Proc. IEEE 51st Vehicular Technology
Conference (VTC 2000-Spring),} vol. 2, pp. 1191-1195, Tokyo,
Japan, May 15-18, 2000.


\bibitem{2003_Sui_CDMAfinger} Haichang Sui, E. Masry, B. D. Rao, Y. C. Yoon, ``CDMA downlink
chip-level MMSE equalization and finger placement," \emph{Proc.
Asilomar Conference on Signals, Systems and Computers,} vol. 1,
pp. 1161-1165, Pacific Grove, CA, Nov. 2003.

\bibitem{2005_Sui_CDMAfinger} Haichang Sui, E. Masry, B. D. Rao, ``RAKE finger
placement for CDMA downlink equalization," \emph{Proc. IEEE
International Conference on Acoustics, Speech, and Signal
Processing,} vol.3, pp. iii/905-iii/908, Philadelphia, PA, March
2005.

\bibitem{2005_Zhiwei_MPfingerSel} Lin Zhiwei, A. B. Premkumar, A. S. Madhukumar,
``Matching pursuit-based tap selection technique for UWB channel
equalization," \emph{IEEE Communications Letters,} vol. 9, pp.
835-837, Sept. 2005.




%\bibitem{eran1} E. Fishler and H. V. Poor, ``On the tradeoff between two types of processing
%gain," \emph{40th Annual Allerton Conference on Communication,
%Control, and Computing,} Monticello, IL, Oct. 2-4, 2002.

\bibitem{eran1} E. Fishler and H. V. Poor, ``On the tradeoff between two types of processing
gain," \emph{IEEE Transactions on Communications,} vol. 53, no.
10, pp. 1744-1753, Oct. 2005.

\bibitem{sinan_WCNC04} S. Gezici, H. Kobayashi, H. V.
Poor and A. F. Molisch, ``Performance evaluation of impulse radio
UWB systems with pulse-based polarity randomization in
asynchronous multiuser environments," \emph{Proc. IEEE Wireless
Communications and Networking Conference (WCNC 2004),} vol. 2, pp.
908-913, Atlanta, GA, March 2004.

\bibitem{paul} Y.-P. Nakache and A. F. Molisch, ``Spectral shape of UWB
signals - Influence of modulation format, multiple access scheme
and pulse shape," \emph{Proc. IEEE 57th Vehicular Technology
Conference, (VTC 2003-Spring)}, vol. 4, pp. 2510-2514, Jeju,
Korea, April 2003.

\bibitem{sinan_T-SP} S. Gezici, H. Kobayashi, H. V. Poor and A. F. Molisch,
``Performance evaluation of impulse radio UWB systems with
pulse-based polarity randomization,'' \emph{IEEE Transactions on
Signal Processing,} vol. 53, no. 7, pp. 2537-2549, July 2005.

\bibitem{sinanUWBST04} S. Gezici, H. Kobayashi, H. V. Poor, and A. F. Molisch,
``Optimal and suboptimal linear receivers for time-hopping impulse
radio systems," \emph{Proc. IEEE Conference on Ultra Wideband
Systems and Technologies (UWBST 2004),} Kyoto, Japan, May 18-21,
2004.

\bibitem{molischWPMC03} A. F. Molisch, Y. P. Nakache, P. Orlik, J. Zhang, Y. Wu, S.
Gezici, S. Y. Kung, H. Kobayashi, H. V. Poor, Y. G. Li, H. Sheng
and A. Haimovich, ``An efficient low-cost time-hopping impulse
radio for high data rate transmission," \emph{Proc. IEEE 6th
International Symposium on Wireless Personal Multimedia
Communications (WPMC 2003),} Yokosuka, Kanagawa, Japan, Oct.
19-22, 2003.

\bibitem{verdu} S. Verd\'{u}. \emph{Multiuser Detection,} Cambridge
University Press, Cambridge, UK, 1998.

\bibitem{optTextbook} S. Boyd and L. Vandenberghe, \emph{Convex
Optimization,} Cambridge University Press, Cambridge, UK, 2004.

\bibitem{Nesterov_IntPointBOOK} Y. Nesterov and A. Nemirovskii, \emph{Interior-Point Polynomial
Methods in Convex Programming,} Society for Industrial and Applied
Mathematics Press, Philadelphia, PA, 1994.

\bibitem{GA1} R. L. Haupt and S. E. Haupt, \emph{Practical Genetic
Algorithms,} John Wiley \& Sons Inc., New York, 1998.

\bibitem{GA2} D. E. Goldberg, \emph{Genetic Algorithms in Search, Optimization, and Machine
Learning,}Addison-Wesley, Reading, MA, 1989.

\bibitem{GA3} M. Mitchell, \emph{An Introduction to Genetic
Algorithms,} MIT Press, Cambridge, MA, 1996.

\bibitem{Juntti_1997} M. J. Juntti, T. Schl\"osser and J. O.
Lilleberg, ``Genetic algorithms for multiuser detection in
synchronous CDMA,'' \emph{Proc. IEEE International Symposium on
Information Theory}, p. 492, Ulm, Germany, June 29-July 4, 1997.

\bibitem{Ergun_2000} C. Erg\"un and K. Hacioglu, ``Multiuser detection
using a genetic algorithm in CDMA communications systems,''
\emph{IEEE Transactions on Communications,} vol. 48, no. 8, pp.
1374-1383, Aug. 2000.

\bibitem{Hanzo_2004} K. Yen and L. Hanzo, ``Genetic-algorithm-assisted
multiuser detection in asynchronous CDMA communications,''
\emph{IEEE Transactions on Vehicular Technology,} vol. 53, no. 5,
pp. 1413-1422, Sept. 2004.


\end{thebibliography}
\end{document}